\documentstyle[12pt]{article}
\def\be{\begin{equation}}
\def\ee{\end{equation}}
\def\ben{$$}
\def\een{$$}
\def\ba{\begin{array}{c}}
\def\ea{\end{array}}
\def\p{\partial}
\begin{document}

\titlepage
\vspace*{2cm}

 \begin{center}{\Large \bf
 ${\cal PT}-$symmetric regularizations in
 supersymmetric quantum mechanics
   }\end{center}

\vspace{10mm}

 \begin{center}
Miloslav Znojil

 \vspace{3mm}

\'{U}stav jadern\'e fyziky AV \v{C}R, 250 68 \v{R}e\v{z}, Czech
Republic\footnote{e-mail: znojil@ujf.cas.cz}

\end{center}

\vspace{5mm}


\section*{Abstract}

Within the supersymmetric quantum mechanics the necessary
regularization of the poles of the superpotentials on the real
line of coordinates $x$ may be most easily mediated by a small
constant shift of this axis into complex plane. Detailed attention
is paid here to the resulting recipe which works, in effect, with
non-Hermitian (a. k. a. ${\cal PT}-$symmetric or pseudo-Hermitian)
Hamiltonians. Besides an exhaustive discussion of the role of the
complex spike in harmonic oscillator, we mention also some
applications concerning the regularized versions of the
Smorodinsky-Winternitz and Calogero models and of the relativistic
Klein-Gordon equation.

\vspace{5mm}

PACS   03.65.Fd; 03.65.Ca; 03.65.Ge; 11.30.Pb; 12.90.Jv

\newpage


 \section{Introduction: Symmetries}

The concept of symmetry proves extremely productive in various
branches of physics. In particular, its applications in quantum
mechanics simplify the analysis of the properties of a symmetric
system characterized by the commutativity of its Hamiltonian
$H_{SY}$ with some operators $S=S_{SY}$ which are tractable as
elements of a Lie algebra of symmetries ${\cal G}_{(SY)}$. Such an
additional information about the Schr\"{o}dinger bound-state
problem simplifies its solution. For example, during the
construction of bound states in a $D-$dimensional central
potential, the use of its symmetry ${\cal G}_{(SY)}=so(D)$ enables
us to reduce the {\em partial} differential Schr\"{o}dinger
equation to its {\em exactly solvable} angular part $
S_{(SY)}\,|n\rangle =\ell(\ell+1)\, |n\rangle$, accompanied by the
{\em ordinary} radial differential equation.

A schematic illustrations of such a method is provided by the
$D=1$ Hamiltonian
 \be
 H_{SY} = -\frac{d^2}{dx^2} + V_{SY}(x), \ \ \ \ \ \ \ \ x \in
 (-\infty,\infty)
 \label{SE1}
 \ee
where the real and confining potential is chosen as spatially
symmetric, $V_{SY}(x) = V_{SY}(-x)$. This means that the
Hamiltonian itself commutes with the operator of the parity $
S_{SY}\equiv {\cal P}$. Each bound state $|n\rangle$ must be an
eigenstate of both the Hamiltonian $H_{SY}$ {\em and} the
operator(s) $S_{SY}$ so that we may select its parity in advance
and, then, evaluate the wavefunction $|n\rangle$ on the mere
half-axis of $x \in (0, \infty)$.

Beyond the similar Lie-algebraic applications, a natural
generalization of the concept of symmetry is encountered in the so
called supersymmetric quantum mechanics (SUSY QM, cf. the review
paper \cite{CKS}) and in the so called ${\cal PT}-$symmetric
quantum mechanics (PTSY QM, cf. its recent formulations in Refs.
\cite{BB,ostatni}).  In these two cases, the linear Lie algebra of
symmetries ${\cal G}_{(SY)}$ is being replaced by a graded Lie
algebra ${\cal G}_{(SUSY)}$ and by its antilinear (i.e., strictly
speaking, non-linear) analogue ${\cal G}_{(PTSY)}$, respectively.
Here we intend to pay attention to the various combinations of
these two possibilities.


In the former context, transition to SUSY QM is based on a deeper
mathematical interpretation of the creation and annihilation
operators. On a less abstract level, SUSY QM works with the
doublets of the [say, ``left" (or $_L-$subscripted) and ``right"
(or $_R-$subscripted)] Hamiltonians arranged in a single
super-Hamiltonian
 \be
 H_{SUSY}=
  \left [ \begin{array}{cc} H_{(L)}&0\\ 0&H_{(R)}
 \ea
 \right ],\ \ \ \ \ \
 H_{(L,R)} = -\frac{d^2}{dx^2} + V_{(L,R)}(x)
 \label{model}
 \ee
which commutes with the two matrix operators which are called
super-charges,
 \be
 S_{SUSY,A} =\left [
 \begin{array}{cc} 0&0\\ A^{}&0
 \ea
 \right ]\equiv{
 \cal Q},
 \ \ \ \ \ \
 S_{SUSY,B}=\left [
 \begin{array}{cc}
0& B^{}
\\
0&0 \ea \right ] \equiv
\tilde{\cal Q}\,.
\label{superch}
 \ee
In 1981, E. Witten \cite{Witten} noticed the relevance of such a
model for clarification of the absence of any SUSY partners of the
observable elementary particles. He tried to attribute their
``invisibility in experiments" to the spontaneous breakdown of
SUSY. Paradoxically, during the last cca twenty years, the idea
found the majority of its applications within non-relativistic
quantum mechanics~\cite{onde,tady}.


The history of the SUSY-inspired search for a symmetry between
bosons and fermions is not too dissimilar to the recent
development of PTSY QM which also finds some of its key physical
motivations within the relativistic quantum field theory
\cite{BBprd}. Many open questions concern the mathematics of PTSY
and, in particular, a deeper understanding of its spontaneous
breakdown \cite{Levai}. Simplified PTSY QM models are studied,
based on the presence of the two real potentials in a {single},
{\em non-Hermitian} Hamiltonian
 \be
 H_{PTSY} = -\frac{d^2}{dx^2} + V_{S}(x)+ i\,V_{A}(x),
  \ \ \ \ \ \ \ \ x \in
 (-\infty,\infty)\,.
 \ee
The two components $V_{S}(x)=V_{S}(-x)$ and $V_{A}(x)=-V_{A}(-x)$
of the complexified force are, by assumption, spatially symmetric
and antisymmetric functions, respectively. The Hamiltonian
commutes, this time, with the product $S_{PTSY}\equiv {\cal
P}\cdot{\cal T}$ of the parity ${\cal P}$ with an auxiliary
antilinear operator ${\cal T}$ which mimics the time
reversal~\cite{BB}.

In a broader context, operator ${\cal P}$ may be generalized to
any Hermitian and invertible ``metric" ${\cal P}={\cal P}^\dagger$
(sometimes assumed to be positive definite \cite{Geyer}) while its
partner ${\cal T}$ mediates Hermitian conjugation \cite{BBJ,BBJb}.
This means that we may eliminate the explicit use of ${\cal T}$
and treat the PTSY-related commutativity
 \be
 S_{PTSY}H_{PTSY}
-H_{PTSY}S_{PTSY}=0
 \ee
as a  mere pseudo-Hermiticity requirement \cite{Dirac}
 \be
 H_{PTSY}^\dagger = {\cal P}\,H_{PTSY}\,{\cal P}^{-1}\,.
 \label{nehe}
 \ee
The best known phenomenological, {\em experimentally} relevant
Hamiltonians with the ``weakened Hermiticity" property
(\ref{nehe}) appear in the Feshbach-Villars version of the
relativistic Klein-Gordon equation (cf. section \ref{relativ}
below). More than ten years ago, pseudo-Hermitian Hamiltonians
also became popular in many phenomenological descriptions of the
bosonic degrees of freedom within atomic nuclei~\cite{Geyer}.

We may summarize that the common Lie-algebraic $S_{SY}$ finds its
close parallels within SUSY QM and PTSY QM.  In the present paper,
we intend to develop, in due detail, several ideas which emerged
in our papers \cite{PLB}-\cite{NPB} where we outlined some merits
of a synchronized work within a {\em combined} ${\cal
PT}-$symmetric {\em and} superymmetric quantum mechanics (PTSUSY
QM).

\section{Conventional regularizations in SUSY QM }

The key SUSY assumption of commutativity may be read,
alternatively, as a factorization postulate for ${ \cal H}$,
 \be
  \left [ \begin{array}{cc} H_{(L)}&0\\ 0&H_{(R)}
 \ea
 \right ]= \left [ \begin{array}{cc} BA&0\\ 0&AB
 \ea
 \right ]\,.
 \label{facto}
 \ee
This postulate implies the further relations
 \ben
 \{ {\cal Q},\tilde{\cal Q}
\}={\cal H} , \ \ \ \ \ \ \{ {\cal Q},{\cal Q} \}= \{ \tilde{\cal
Q},\tilde{\cal Q} \}=0, \ \ \ \ \ \ \ \ [ {\cal H},{\cal Q} ]=[
{\cal H},\tilde{\cal Q} ]=0\,
  \een
which, in their turn, enable us to treat the supercharges
$S_{SUSY}$ as generators of the graded Lie algebra sl(1/1).

Example (\ref{SE1}) with the $D-$dimensional harmonic-oscillator
$V_{(SY)}^{(HO)}(|\vec{r}|)= |\vec{r}|^2$ offers one of the most
transparent illustrations of the essence of many of the
characteristic ideas of SUSY QM.  Thus, at $D=1$ one inserts the
two differential operators $A = \p_x+x$ and $B = -\p_x+x$ of the
first order in the explicit definition (\ref{superch}) of the two
supercharge operators.  The resulting explicit SUSY Hamiltonian
(\ref{model}) connects the two {\em regular} harmonic oscillators
$ H_{(L,R)}^{(HO)} = -{d^2}/{dx^2} + x^2 \mp 1$. The wave
functions pertaining to these oscillators are analytic and may
serve, in parallel, as an illustration of a transition to the
complexified coordinates $x$ in PTSY QM of Ref.~\cite{BB}.


After a transition to $D>1$, equation (\ref{SE1}) remains almost
the same, complemented merely by a strongly singular kinetic term
in $V_{SY}(x)\to V_{SY}(x) +V_{kinetic}(x)$ where $V_{kinetic}(x)
=\ell(\ell+1)/x^2 $ with $\ell =\ell(n)= (D=3)/2+n$ and
$n=0,1,\ldots$ \cite{BG}. Marginally, it is rather amusing to
notice that precisely the difference between the absence and
presence of the strongly singular $V_{kinetic}(x)$ plays a key
role in SUSY QM. Indeed, one of the characteristic features of the
formalism of SUSY QM is the requirement of the absence of the real
poles in its partner potentials $ V_{(L,R)}$ (see, e.g., the
review paper \cite{CKS} for more details). {\em Vice versa}, a
puzzling difficulty is immediately encountered when we try to
study a SUSY model with such a singularity. During the detailed
analysis of the central harmonic oscillator (HO) in $D>1$
dimensions, Jevicki and Rodrigues (JR, \cite{JR}) revealed that in
such a singular case, the textbook recipes lead to a {\em
singular} form of the HO superpotential,
 \be
 W(x) = W^{(\gamma)}(x) = x - \frac{\gamma+1/2}{x}
 \label{suwi}
 \ee
where $\gamma$ may vary with $D$ etc \cite{hoptsusy}. Then, at any
real strength parameter $\gamma$, the naive SUSY algebra is well
known to generate the formal partnership between the two harmonic
oscillator Hamiltonian operators of the form $H^{(\kappa)} =
-d^2/dx^2 + x^2+(\kappa^2-1/4)/x^2$ where $\kappa = \kappa_{L,R} =
\ell_{L,R}+1/2$ \cite{CKS}. Of course, in order to satisfy the two
factorization requirements in eq. (\ref{facto}) we must define
 \be
 H_{(L)} = H^{(\alpha)} - 2\gamma-2, \ \ \ \ \ \ \
 H_{(R)} = H^{(\beta)} - 2\gamma\,
 \label{spektruf}
 \ee
and put $\alpha = |\gamma| = \ell_{(L)}+1/2$ and $\beta =
|\gamma+1| = \ell_{(R)}+1/2$. This means that people usually
restrict their attention to the regular or ``linear" HO
Hamiltonians $H^{(\pm 1/2)}$ defined on the whole axis of $x \in
(-\infty,\infty)$ and possessing the well known energy spectrum
 \be
 E^{(LHO)}_n = 2n+1, \ \ \ \ n = 0, 1, \ldots\,.
 \ee
In such a regular case (i.e., at the exceptional $\gamma=-1/2$)
the partner SUSY spectra $[E^{(\gamma)}_{(L,R)}]_n$ reflect the
shifts in eq. (\ref{spektruf}) and exhibit the expected SUSY-type
isospectrality,
 \be
 E_{(L)}^{(-1/2)} = \{ 0, 2, 4, 6, 8, \ldots \},\ \ \ \ \ \
 E_{(R)}^{(-1/2)} = \{ 2, 4, 6, 8, \ldots \}\,.
 \ee
More rarely, one may arrive at the similar conclusions using the
{\em more general} or ``singular" HO Hamiltonian operators
$H^{(\kappa)}$ which {\em must be restricted} to the {smaller}
domain of $x \in (0,\infty)$ [or, if necessary, $x \in
(-\infty,0)$] and possess, therefore, the {\em different} energy
spectrum $E^{(SHO)}_n = 4n+2\ell+3$.

On this background it is easy to characterize the essence of the
puzzle described by Jevicki and Rodrigues \cite{JR} as an
attempted study of the relationship between the {\em same}
operators defined on the {\em different} domains (see Ref.
\cite{hoptsusy} for more details). For illustration of the essence
of this puzzle, it is most popular to pick up the $\gamma\neq
-1/2$ superpotential (\ref{suwi}), say, with $\gamma=1/2$. Then,
the constant shifts in eq. (\ref{spektruf}) give us immediately
the very different ``partner" spectra
 \be
 E_{(L)}^{(1/2)} = \{ -2, 0, 2, 4, 6, 8, \ldots \}, \ \ \ \ \
 E_{(R)}^{(1/2)}
 = \{  4,  8, 12, 16, \ldots \}\,.
 \ee
The explanation of the sudden loss of the desired SUSY-type
isospectrality lies in the {\em one-sided} emergence of the
centrifugal-like singularity $2/x^2$ in the right SUSY partner
$H_{(R)}$ only. At the same time, the left operator $H_{(L)}$
remains regular on the whole real line. The complete set of the
left normalizable wave functions is not constrained by the
boundary condition applied to the right wavefunctions in the
origin,  $\psi_{(R)}(0)=0$. This observation might inspire a naive
resolution of the JR paradox, based on an addition of the
artificial and arbitrary left boundary condition
$\psi_{(L)}(0)=0$. This would reduce the left spectrum to the set
$ E_{(L)}=\{0,  4, 8, \ldots \}$ which obeys the required SUSY
isospectrality rule.

An alternative, less naive though much more cumbersome
regularization recipe has been offered by Das and Pernice
\cite{DP} who re-defined slightly $H_{(L)}$ and $H_{(R)}$ via an
additional delta function in the origin. In effect, their
construction proves equivalent~\cite{NPB} to the use of the
Dirichlet boundary condition $\psi_{(R)}(0)=0$ for the right
Hamiltonian {\em in a rather unusual combination} with the Neumann
left boundary condition, $\p_x\psi_{(L)}(0)=0$. They obtained
 \be
 E_{(L)}^{(1/2)} = \{ 0, 2, 4, 6, 8, \ldots \},\ \ \ \ \ \
 E_{(R)}^{(1/2)} = \{ 2, 4, 6, 8, \ldots \}\,.
 \ee
In our present paper we intend to emphasize that a much easier and
more natural approach to the resolution of the JR regularization
puzzle may be based on a downward complex shift of the real axis
of $x$. Besides giving an outline of its essence (i.e., of an
analytic PTSY extension of quantum mechanics), we shall also
mention some of the possible consequences of this approach for a
few other important SUSY-related models with singularities.


 \section{Regularized constructions within PTSUSY QM }

\subsection{An exactly solvable model \label{2} \label{2.2}}

Harmonic-oscillator Hamiltonians
 \ben
  H^{(\alpha)}= -
 \frac{d^2}{dr^2}
+
 \frac{\alpha^2-1/4}{r^2}
+ r^2, \ \ \ \ \ \ \alpha > 0\  \label{nasham}
 \een
with the Buslaev's and Grecchi's \cite{BG} PTSY regularization of
$r = x - i\,\varepsilon$ (with real $x$, cf. also Ref.
\cite{Ahmed} in this context) have thoroughly been studied in Ref.
\cite{ptho}. Their normalizable wave functions
 \be
 \langle r | N ^{(\varrho)} \rangle =
\frac{N!}{\Gamma(N+\varrho+1)}\cdot
 r^{\varrho+1/2} \exp(-r^2/2) \cdot L_N^{(\varrho)}(r^2)
 \label{hu}
  \ee
and energies
  \ben
  E_{}^{}
  = E_{N}^{(\varrho)} =4N+2\varrho+2,
\ \ \ \ \ \ \ \
 \varrho= -Q \cdot
\alpha\
  \een
were described there as manifestly dependent on the quantum number
$Q = \pm 1$ of the so called quasi-parity. In the context of
fields, this quantity proves intimately related to the more recent
concept of the so called charge-conjugation symmetry ${\cal C}$ as
introduced, e.g., in  Ref.~\cite{BBJ}.

We already mentioned that the regularized superpotentials
$W^{(\gamma)}$ enter the SUSY operators
$A^{(\gamma)}=\p_q+W^{(\gamma)}$ and
$B^{(\gamma)}=-\p_q+W^{(\gamma)}$ in a way which defines the
``left" and ``right" components of the super-Hamiltonian ${\cal
H}$,
 \ben
 H_{(L)}=B\cdot A=\hat{p}^2+W^2-W'
 , \ \ \ \ \ \ \ \
 H_{(R)}=A \cdot B=\hat{p}^2+W^2+W'.
 \een
For our present choice of $W=W^{(\gamma)}$ these SUSY partners are
$\gamma-$dependent and, generically (i.e., at any complex
$\gamma$), both of them are proportional to the {\em different}
harmonic oscillators,
 \be
 {H}_{(L)}^{(\gamma)} = {H}_{}^{(\alpha)} -2\gamma-2,
 \ \ \ \ \ \
 {H}_{(R)}^{(\gamma)} = {H}_{}^{(\beta)} -2\gamma, \ \ \ \
 \ \ \ {\alpha}^2=\gamma^2, \ \ \ \
 \ \ \ \beta^2=(\gamma+1)^2\
 \label{Mtt}
 \ee
with, conventionally, ${\rm Re}\ \alpha> 0$ and ${\rm Re}\ \beta>
0$.

\subsection{SUSY under unbroken ${\cal PT}$ symmetry \label{2.2.1}}

At all the real $\gamma \in (-\infty,\infty)$ we select $ \alpha>
0$ and $\beta> 0$. This means that all the energies remain real
and that one may distinguish between the following three regimes,
 \be
 \left \{
 \begin{array}{ll}
 {\rm positive}\ \gamma = \alpha > 0,
& {\rm dominant} \   \beta = \alpha + 1,
 \\
{\rm small\ negative}\ \gamma = - \alpha > -1, &   {\rm
comparability,}
   \   \alpha + \beta =1, \\
{\rm large\ negative}\ \gamma = - \alpha < -1, &  {\rm dominant} \
 \alpha = \beta + 1.
 \ea
 \right .
  \ee
For a given ``left" Hamiltonian $H_{(L)} \sim H^{(\alpha)}$ we
have a choice between $\gamma = \pm \alpha$ giving the two
alternative ``right" partners $H^{(\beta_{1,2})}$ such that
$\beta_{1,2} = 1 \pm \alpha$ for small $\alpha < 1/2$ while
$\beta_1 = |\alpha - 1| < \alpha < \beta_2 = \alpha + 1$ for
$\alpha > 1/2$. Each pair of the SUSY partners
${H}_{(L,R)}^{(\gamma)}$ generates the quasi-even and quasi-odd
energies and these energies form an ordered quadruplet at any main
quantum number $n$,
 \be
 \left \{
 \begin{array}{ll}
 E_{(L)}^{(-\alpha)} \ =
 E_{(R)}^{(-\beta)} \  <
 E_{(L)}^{(+\alpha)} \  <
 E_{(R)}^{(+\beta)} \ & {\rm for \ positive}\ \gamma ,
 \\
 E_{(L)}^{(-\alpha)} \ <
 E_{(R)}^{(-\beta)} \  =
 E_{(L)}^{(+\alpha)} \  <
 E_{(R)}^{(+\beta)} \ & {\rm for \ negative}\ \gamma  > -1,
\\
 E_{(L)}^{(-\alpha)} \ <
 E_{(R)}^{(-\beta)} \  <
 E_{(L)}^{(+\alpha)} \  =
 E_{(R)}^{(+\beta)} \ & {\rm for \ negative}\ \gamma  < -1
 \ea
 \right .
  \ee
(mind and mend the misprints in Table 1 of Ref. \cite{hoptsusy}
and consult also Figure 1 {\it ibidem}). The equal-$n$
degeneracies may be emphasized by the square brackets ``[" and
``]",
 \be
 \left \{
 \begin{array}{ll}
[4n-4\alpha ] < 4n < 4n+4 \ & {\rm for \ positive}\ \gamma ,
 \\
 4n < [ 4n+4\alpha] < 4n+4\ & {\rm for \ negative}\ \gamma  > -1,
\\
 4n<4n+4 < [4n+4\alpha] \ & {\rm for \ negative}\ \gamma  < -1.
 \ea
 \right .
  \ee
By their (linear) $\alpha-$dependence, they are distinguished from
the usual non-equal-$n$ degeneracies with the values of energies
which do not vary with this parameter.

 \subsection{Spontaneously broken symmetry regime\label{2.2.2}}

The ${\cal PT}$ symmetry of the ``left" Hamiltonians $H_{(L)}\sim
H^{(\alpha)}$ becomes spontaneously broken at the purely imaginary
$\alpha $~\cite{ptho}. This results from the purely imaginary
choice of $\gamma $ in the superpotential $W^{(\gamma)}$. We have
to distinguish between the two separate half-lines,
 \be
 \left \{
 \begin{array}{lll}
\delta > 0 \ {\rm in}\ \gamma =i\,\delta, & \alpha = i\,\delta,\ &
\beta = 1+\alpha,
 \\
\eta > 0 \  {\rm in}\ \gamma =-i\,\eta, \ & \alpha = i\,\eta,\ &
\beta =  1- \alpha.
 \ea
 \right .
  \ee
On both of them the energies pertaining to the unshifted and
${\cal PT}$ symmetric Hamiltonian $H^{(\alpha)}$  occur in the
complex conjugate pairs  $E_n^{(\pm \alpha)}$ \cite{Mostafazadeh}.
This symmetry is broken by the shift $H^{(\alpha)}\to
H^{(\gamma)}_{(L)}$ enforced by the SUSY factorization in eq.
(\ref{Mtt}). Hence, the supersymmetrization still generates the
partially real spectra even at the purely imaginary $\alpha =
\sqrt{\gamma^2}$,
 \ben
 \left \{
 \begin{array}{llll}
 E_{(L)}^{(+ \alpha)} = 4n,\
&  E_{(L)}^{(- \alpha)} = E_{(R)}^{(- \beta)} = 4n-4\alpha,\ &
 E_{(R)}^{(+ \beta)} = 4n+4\ &
{\rm for }\  \gamma = i\,\delta,
 \\
 E_{(L)}^{(- \alpha)} = 4n,\ & E_{(L)}^{(+ \alpha)} =
  E_{(R)}^{(- \beta)} =
 4n+4\alpha,\
& E_{(R)}^{(+ \beta)} =
 4n+4\
&
  {\rm for }\  \gamma = -i\,\eta .
 \ea
 \right .
  \een
We see that the imaginary components either vanish or are equal to
$4\gamma$. In the latter case the complex ``left" and ``right"
energies coincide at a fixed $n$. The change of the sign of
$\gamma$ causes their complex conjugation. Simultaneously, the
real parts of the energies are $\gamma-$independent and
equidistant, and the characteristic degeneracy pattern $E_{(L)}(n)
= E_{(R)}(n-1)$ holds for all the {\em real} levels in the
spectrum exempting, as usual, the non-degenerate $n=0$ state.

 \subsection{Completely broken symmetry regime
  \label{2.2.3}}

We have seen that the spontaneous breakdown of the ${\cal PT}$
symmetry of our initial or ``left" Hamiltonian $H_{(L)}$ leads to
the mere {\em partial} complexification of the related energies
$E_{(L)}$. One may reformulate this observation by saying that the
SUSY construction keeps some ``right" energies $E_{(R)}$ real even
though the ${\cal PT}$ symmetry of the ``right" Hamiltonian
$H_{(R)}$ itself is violated in the {\em manifest} manner.

The latter empirical rule may be easily generalized by induction.
For this purpose let us choose a positive integer $N$ and start
from the superpotential $W^{(\gamma)}$ with a complex parameter
$\gamma = N+i\,q\,\delta$ with $\delta > 0$ and $q = \pm 1$. In
the first step we get the SUSY rules $\alpha^2 = \gamma^2$ and
$\beta^2=(\gamma+1)^2$ and make our $\alpha$ and $\beta$ unique by
the constraints ${\rm Re}\ \alpha > 0$,  ${\rm Im}\ \alpha =
\delta> 0$ and ${\rm Re}\ \beta > 0$. This gives the two options,
 \ben
 \beta_1 = N-1+i\,\delta < \alpha = N + i\,\delta < \beta_2 =
 N+1+i\,\delta
  \een
i.e., either a return to the previous $N$ or the induction step
towards the next one. Now we may summarize that at all these $N$
the SUSY rules lead, up to the ambiguity in signs, to the same
conclusion as above, giving
 \be
 E_{(L)} = \left \{
 \begin{array}{ll}
 4n-4\gamma
 \\
 4n
 \ea
 \right . , \ \ \ \ \
 E_{(R)} = \left \{
 \begin{array}{ll}
 4n-4\gamma
 \\
 4n+4
 \ea
 \right . .
 \label{partially}
  \ee
In this way, the picture is closed and complete: The class of the
spiked harmonic oscillator SUSY partners possessing the same
coinciding and partially real spectra (\ref{partially}) may be all
derived from the superpotential $W^{(\gamma)}$ defined on the
lattice of $\gamma = N \pm i\,\delta$ with integers $N = 0, 1,
\ldots $ and positive $\delta > 0$. The $N=0$ observations of
subsection \ref{2.2.2} remain valid at all $N$. Also the ${\cal
PT}$ symmetric limit $\delta \to 0$ moves us to the exceptional
points $\alpha = {\rm integer}$ where one encounters the
Jordan-block structures and unavoided crossings of the energy
levels~\cite{Herbst,ptho,AM}.

\section{An alternative supersymmetrization  \label{3}}

The most striking feature of the above SUSY scheme lies in the
puzzling jumps between the different Hamiltonians $H^{(\alpha)}$
and $H^{(\beta)}$. The action of the linear differential operators
$A^{(\gamma)}=\p_q+W^{(\gamma)}$ and
$B^{(\gamma)}=-\p_q+W^{(\gamma)}$ on the states
$|n^{(\gamma)}\rangle$ may change both their quantum number $n$
{\em and} the parameter $\gamma$.

A transition from $\gamma$ to quasi-parity $Q$ and/or to the two
``physical" spike-strength parameters $\alpha $ and $\beta$
clarifies that the ``natural" connection between the SUSY and
annihilation and creation operators remains internally consistent
in the exceptional $\alpha \to 1/2$ limit only. Fortunately, in
all the other cases, a return $\beta \to \alpha$ to the original
system may still be mediated by elementary operators. The
determination of their action on the Laguerre polynomials
appearing in the explicit form of our present wave functions is
easy and the rule we need is given by the quadratic operators
 \ben
  {\bf B} (\alpha)=
B^{(-\gamma)} \cdot B^{(\gamma-1)}\,\equiv\, B^{(\gamma)} \cdot
B^{(-\gamma-1)}, \ \ \ \ \ \ \ \ \ {\bf A}(\alpha)=
A^{(-\gamma-1)} \cdot A^{(\gamma)} \,\equiv\, A^{(\gamma-1)} \cdot
A^{(-\gamma)}
 \een
of Ref.~\cite{hoptsusy} (cf. also \cite{Andrianov} and
\cite{Quesne}) which satisfy the relations
  \ben
{\bf B}(\alpha) \,
 \left |N^{(\gamma)} \right \rangle=c(N,\gamma)\,
  \left |(N+1)^{(\gamma)} \right \rangle,\ \ \ \ \ \ \ \
 {\bf A}(\alpha) \,
   \left |(N+1)^{(\gamma)} \right \rangle=c(N,\gamma)\,
   \left |N^{(\gamma)} \right \rangle
  \een
where $\gamma=\pm \alpha $ may now be complex while the formula
$c(N,\gamma)=-4\sqrt{(N+1)(N+\gamma+1)}$ of Ref.~\cite{hoptsusy}
remains unchanged.

The latter equations may be interpreted as the respective creation
and annihilation of a state at a quasi-parity $Q = -{\rm sign\
(Re}\ \gamma)$. One may, if necessary, exempt here the anomalous
integers $ \alpha = \alpha_c=0, 1, 2, \ldots$ at which one of the
coefficients $c(N,\gamma)$ vanishes and where the two different
energy levels cross or, in other words, where the Bender-Wu
singularities \cite{BW} become real.


A possible deeper meaning of the peculiar second-order
differential operators $ {\bf B}(\alpha)$ and $ {\bf A}(\alpha)$
is to be sought now by the straightforward calculation which
reveals that
 \ben
  {\bf A}(\alpha) \,
   {\bf B}(\alpha) \,-\, {\bf B}(\alpha) \,
    {\bf A}(\alpha) \,\equiv\, 8\,H^{(\alpha)}.
    \een
This means that in spite of their unusual form, our innovated
creation and annihilation operators exhibit an unexpected
simplicity. A ``hidden" algebraic meaning of the above rule may be
significantly clarified by the explicit evaluation the two further
commutators,
 \ben
  {\bf A}(\alpha) \,H^{(\alpha)}\,-
   \,H^{(\alpha)}\, {\bf A}(\alpha)
   \,\equiv\,4\, {\bf A}(\alpha), \ \ \ \ \ \
  \,
   \,H^{(\alpha)}\,{\bf B}(\alpha)\, -\,
   {\bf B}(\alpha) \,H^{(\alpha)} \,\equiv\,4\, {\bf B}(\alpha).
    \een
This implies that the three differential operators $ {\bf
A}(\alpha)/\sqrt{32}$, ${\bf B}(\alpha)/\sqrt{32}$ and
$H^{(\alpha)}/4$ of the second order are generators of the Lie
algebra $sl(2,I\!\!\! R)$.


By the existence of the symmetries just revealed, we feel
encouraged to define the new set of operators
 \be
 {
 \bf G}= \left [ \begin{array}{cc} {\bf G}_{(L)}&0\\ 0&{\bf G}_{(R)}
 \ea
 \right ]
, \ \ \ \ \ \
 {
 \bf Q}=\left [
 \begin{array}{cc} 0&0\\ {\bf A}(\alpha)&0
 \ea
 \right ],
 \ \ \ \ \ \
\tilde{\bf Q}=\left [
 \begin{array}{cc}
0&  {\bf B}(\alpha)
\\
0&0 \ea \right ]\label{generatorsd}
 \ee
which may be perceived as a modified or alternative SUSY recipe.
Once we put
 \be
 {\bf G}_{(L)} ={\bf B}(\alpha) \,{\bf A}(\alpha) \,,\ \ \ \ \ \
 \ \ \
 {\bf G}_{(R)}={\bf A}(\alpha) \,{\bf B}(\alpha) \,
 \label{multi}
 \ee
the new operators (\ref{generatorsd}) generate a representation of
the SUSY algebra sl(1/1),
 \ben
 \{ {\bf Q},\tilde{\bf Q}
\}={\bf G} , \ \ \ \ \ \ \{ {\bf Q},{\bf Q} \}= \{ \tilde{\bf
Q},\tilde{\bf Q} \}=0, \ \ \ \ \ \ \ \ [ {\bf G},{\bf Q} ]=[ {\bf
G},\tilde{\bf Q} ]=0.
  \een
The elements of the innovated super-Hamiltonian-like two-by-two
matrix $ {\bf G}$ are differential operators of the fourth order.
They are defined, in the most compact form, by the products
(\ref{multi}) of the two boldface and mutually non-adjoint factors
 \ben
{\bf B}(\alpha)=
  \frac{d^2}{d\,x^2} -
   \frac{\alpha^2-1/4}{(x-i\,\varepsilon)^2}
    +(x-i\,\varepsilon)^2-
 2\, (x-i\,\varepsilon)\, \frac{d}{d\,x} -1, \een
 \ben
 {\bf A}(\alpha)
 =
  \frac{d^2}{d\,x^2} -
   \frac{\alpha^2-1/4}{(x-i\,\varepsilon)^2}
    +(x-i\,\varepsilon)^2+
 2\, (x-i\,\varepsilon)\, \frac{d}{d\,x} +1.
 \een
The action of both the components of ${\bf G}$ on our basis states
is transparent,
 \ben
 {\bf G}_{(L)}
 \,   \left |N^{(\gamma)} \right \rangle
= \Omega_N^{(\gamma)} \, \left |N^{(\gamma)} \right \rangle\, , \
\ \ \ \ \ \ \ \ \Omega_N^{(\gamma)} =16\,N\,(N+\gamma)\, \een \ben
 {\bf G}_{(R)}
\,
 \left |N^{(\gamma)} \right \rangle=
\Omega_{N+1}^{(\gamma)} \,
 \left |N^{(\gamma)} \right \rangle\,.
 \een
Tractable as the respective creation and annihilation operators,
both $ {\bf B}(\alpha)$ and ${\bf A}(\alpha)$ conserve the
quasi-parity $Q$ (i.e., the sign of the superscript $\gamma$) and
enable us to split the Hilbert space into two separate ``halves"
which may be marked by $Q = \pm 1$.

\section{Towards realistic applications}

One of the key messages of our present paper is its emphasis on
the productivity of a combination of the algebraic ideas of
supersymmetry [illustrated by the factorization (\ref{facto}) of
some elementary harmonic-oscillator-type differential operators]
with the ideas of ${\cal PT}-$symmetry and analytic continuation
(best exemplified by the singularity paradoxes and their
comparatively easy analytic regularizations in complex plane).
Several merits of such a combined approach may be emphasized.
First of all, one becomes able to reveal an algebraic background
of the analytic properties (e.g., an analyticity-based complex
shift of a complex coordinate found its algebraic pseudo-Hermitian
explanation in Refs. \cite{Ahmed,Mostafazadeh}) and vice versa
(for example, the algebraic ``spectral-equivalence" effect of the
same shift has found its analytic-continuation explanation in
Refs. \cite{BB,BG} or \cite{ptho}).

Of course, our natural ambition of extending the similar parallels
beyond the domain of the most elementary illustrative examples is
limited by the numerous technical obstacles. In the second half of
our paper let us pay attention to their selected sample.

\subsection{Angular Schr\"{o}dinger equations in PTSUSY QM}

In the introductory part of the review \cite{CKS} (cf. the picture
Nr. 2.2 there) it has been emphasized that the free motion within
a square well generates the whole SUSY family of the exactly
solvable models, the first nontrivial element of which is the
single-well P\"{o}schl-Teller problem on a finite interval,
 \be
 H_{1W} = -\frac{d^2}{d\varphi^2} + V_{1W}(\varphi)
 , \ \ \ \ \ \ \ \
 V_{1W}(\varphi)=\frac{A(A+1)}{\sin^2\varphi}, \ \ \ \ \ \ \ \
 \varphi \in  (0,\pi)\,.
 \label{SEPoTe}
 \ee
Bagchi et al \cite{sqw} have emphasized that the same observation
remains valid when we consider the PTSY generalization of this
model.

It has been noticed in our recent paper \cite{angulSE} that the
merely slightly modified angular differential equation
 \be
 H_{4W} = -\frac{d^2}{d\varphi^2} + V_{4W}(\varphi)
 , \ \ \ \ \ \ \ \
 V_{4W}(\varphi)=\frac{A(A+1)}{\sin^24\,\varphi}, \ \ \ \ \ \ \ \
 \varphi \in  (0,\pi/4)\,
 \label{QWSEPoTe}
 \ee
is a core of the construction of the bound states in the
non-central potentials
  \be
V^{(SI)}(X,Y)= X^2 + Y^2 +\frac{G}{X^2} +\frac{G}{Y^2}
 \,
 \label{parc} \label{Wint}
  \ee
of the Smorodinnsky-Winternitz ``superintegrable" family in two
dimensions \cite{SmoWin}. In parallel, the famous Calogero's
\cite{Calogero} three-body bound-state model represented by the
separable partial differential equation
   \ben
    \left [ -\frac{\p^2}{\p
X^2} -\frac{\p^2}{\p Y^2}+X^2+Y^2+ \right .
  \een
   \be
 \left .
  +\frac{g}{2\,X^{2}}+\frac{g}{2\,(X-\sqrt{3}Y)^{2}}
+\frac{g}{2\,(X+\sqrt{3}Y)^{2}}
 \right ]
\,\Phi(X,Y)={\cal E} \,\Phi(X,Y) \, \label{Calo}
  \ee
has a very similar ordinary differential angular part,
 \be
 H_{6W} = -\frac{d^2}{d\varphi^2} + V_{6W}(\varphi)
 , \ \ \ \ \ \ \ \
 V_{6W}(\varphi)=\frac{A(A+1)}{\sin^26\,\varphi}, \ \ \ \ \ \ \ \
 \varphi \in  (0,\pi/6)\,.
 \label{seSEPoTe}
 \ee
Unfortunately, the ${\cal PT}-$symmetrization of the latter two
models  does not lead to any simplifications similar to those
which we witnessed, e.g., in section \ref{2} above. At this point,
it is our sad duty to say that an opposite observation is true.
One of the main reasons is that in the Hermitian version of both
the above-mentioned models (where the P\"{o}schl-Teller force
appears in the angular part of the Schr\"{o}dinger equation),
Dirichlet boundary conditions are imposed at both the ends of the
interval of the angle $\varphi$. This reflects the presence of
strongly repulsive barriers at both these exceptional points.

After the PTSY regularization of the coordinate we arrive at the
different family of eigenvalue problems,
 \be
 H_{MW} = -\frac{d^2}{d\varphi^2} + V_{MW}(\varphi)
 , \ \ \ \ \ \ \ \
 V_{MW}(\varphi)=\frac{A(A+1)}{\sin^2M\,\varphi},
 \ee
 \ben
  \ \ \ \ \ M = 1, 2, \ldots,\ \ \ \ \ \ \ \
 \varphi = \alpha - i\,\varepsilon,\ \ \ \ \ \ \ \
 \alpha \in  (0,\pi)
 \,.
 \label{QWSEPoTe}
 \een
Our choice of the complex shift $\varepsilon > 0$ opens the
possibility of the tunnelling through the barriers (cf. the
particular $M=4$ and $M=6$ results in Refs. \cite{Jakubsky} and
\cite{Tater}, respectively). This means that we must impose the
different, periodic boundary conditions at the endpoints of our
domain of the angular coordinate $\alpha$ which is {$M$ times
bigger}.

The phenomenon of the tunnelling between the neighboring
(sometimes called ``Weyl's") chambers is, in this context, a
source of a new physics as well as of a significantly worsened
mathematical insight in the structure of the PTSY solutions at any
$M\geq 2$. This has been illustrated on a schematic semi-numerical
double-well model in Ref. \cite{angulSE}.

\subsection{Relativistic PTSUSY quantum mechanics? \label{relativ}}

Once we moved to the non-Hermitian Hamiltonians, the scope of
quantum mechanics (including also its specific regularization
aspects) finds a fairly natural extension to the domains with
relativistic kinematics.

In order to be more specific, let us pick up the well known
example of the the Klein-Gordon equation in the units $ \hbar = c
=1$,
 \be
 \
 \left( i\,\p_t\right )^2 \Psi^{(KG)}(x,t)
 = H^{(KG)}\,\Psi^{(KG)}(x,t)\,.
 \label{jdvar}
 \ee
This equation is not too different from its Schr\"{o}dinger
non-relativistic predecessor
 \be
 \
 i\,\p_t \Psi^{(NR)}(x,t) = H^{(NR)}\,\Psi^{(NR)}(x,t)\,.
 \label{jednar}
 \ee
After we abbreviate $\varphi_1= i\,\p_t\,\Psi^{(KG)}(x,t)$ and
$\varphi_2= \Psi^{(KG)}(x,t)$, the relativistic evolution may be
described by the equivalent Feshbach-Villars \cite{FV} first-order
differential equation presented here in the form
 \be
 i\,
 \p_t
 \left (
 \ba
 \varphi_1\\ \varphi_2
 \ea
 \right )
=
 \left (
 \begin{array}{cc}
 0&H^{(KG)}\\
 1&0
 \ea
 \right )
 \cdot \left (
 \ba
 \varphi_1\\ \varphi_2
 \ea
 \right )
 \,.
 \label{jednar}
 \ee
In contrast to the freedom of option between Hermiticity and
pseudo-Hermiticity for the current non-matrix Hamiltonians,
 \be
 \left [H^{(NR,KG)}\right ]^\dagger = \eta\,H^{(NR,KG)}\,\eta^{-1},
 \ \ \ \ \ \ \eta = \eta^\dagger\,,
 \label{etaf}
 \ee
the two-by-two Feshbach-Villars matrix operator and equation prove
manifestly non-Hermitian,
 \be
 \left [H^{(FV)} \right ]^\dagger = {\cal P}\,H^{(FV)}\,{\cal
 P}^{-1}, \ \ \ \ \ \ \ \
 H^{(FV)}=\left (
 \begin{array}{cc}
 0&H^{(KG)}\\
 1&0
 \ea
 \right ), \ \ \ \ \ \ \ \
 {\cal P}=\left (
 \begin{array}{cc}
 0&\eta\\
 \eta&0
 \ea
 \right )
 \,.
 \ee
We see that we have to pay the price for the transition from the
nonrelativistic eq. (\ref{jednar}) to its relativistic descendant
(\ref{jdvar}). Irrespectively of whether the initial scalar
operator $H^{(KG)}$ itself is Hermitian, quasi-Hermitian or
pseudo-Hermitian (in correspondence to the respective ``metric"
$\eta=I$, $\eta> 0$ or merely non-singular in eq. (\ref{etaf})),
its new Feshbach-Villars representation $H^{(FV)}$ remains
non-Hermitian.

One may reverse this argument as follows. Once we discover the
merits of a regularization (e.g., of any centrifugal-like
singularity in $H^{(KG)}$) by its (e.g., constant-shift)
complexification, an extension of our previous results to the new
relativistic context becomes almost straightforward. In spite of
the fact that any detailed analysis in this direction would
already lie beyond the scope of our present paper, we must
emphasize that the auxiliary diagonalization of $H^{(FV)}$,
 \be
 \left (
 \begin{array}{cc}
 0&H^{(KG)}\\
 1&0
 \ea
 \right )
 \cdot \left (
 \ba
 |n_1^{(\pm)}\rangle\\ |n_2^{(\pm)}\rangle
 \ea
 \right )=
 E_n^{(\pm)}\,
 \left (
 \ba
 |n_1^{(\pm)}\rangle\\ |n_2^{(\pm)}\rangle
 \ea
 \right )
 \,
 \label{jednadfr}
 \ee
remains feasible and facilitated by its partitioning which means
that
 \be
 H^{(KG)}\,
 |n_2^{(\pm)}\rangle =
 \left [E_n^{(\pm)}\right ]^2 |n_2^{(\pm)}\rangle,
 \ \ \ \ \ \
 |n_1^{(\pm)}\rangle=E_n^{(\pm)}\, |n_2^{(\pm)}\rangle
 \,.
 \label{jeddfr}
 \ee
Hence, for all the positive operators $H^{(KG)}>0$ and for all the
present SUSY and/or PTSY regularization purposes we may work with
the modified Schr\"{o}dinger-type equation (\ref{jeddfr}), knowing
that the superscript $^{(\pm)}$ indicates merely the sign of the
real ``energy" $E_n^{(\pm)} \equiv \pm |E_n^{(\pm)}|$. In
parallel, one still has to keep in mind that even for all the
Hermitian $H^{(KG)}$ with $\eta=I$, the FV-time-evolution itself
remains pseudo-unitary \cite{ego} since
 \be
 {\cal P}=\left (
 \begin{array}{cc}
 0&I\\
 I&0
 \ea
 \right )=
 \left [
  \frac{1}{\sqrt{2}}\,
 \left (
 \begin{array}{cc}
 I&I\\
 I&-I
 \ea
 \right )
 \right ]\,\left (
 \begin{array}{cc}
 I&0\\
 0&-I
 \ea
 \right )\,
 \left [
  \frac{1}{\sqrt{2}}\,
 \left (
 \begin{array}{cc}
 I&I\\
 I&-I
 \ea
 \right )
 \right ]
 \,.
 \ee
Thus, the quantized Klein-Gordon system may be used and presented
as an archetypal physical PTSY model \cite{pseudo,CzW}, weakening
the above limitation of our scope to the mere non-relativistic
cases.

\section{Concluding remarks and open questions}

Let us summarize that without a transition to the language of PTSY
QM, virtually all the central oscillators have a perceivably
different mathematical interpretation at $D=1$ (in one dimension)
and at $D> 1$ (when there emerges a strongly singular centrifugal
force in eq. (\ref{SE1})). After this transition, the mathematical
difference disappears with the re-emergence of the safely
normalizable bound states of an even quasi-parity. In the new
language, it is less puzzling and purely technical to select and
characterize the unphysical character of all the quasi-even states
at all the higher spatial dimensions $ D \neq 1$.

We reminded the readers that the difference between the present
and vanishing centrifugal singularity (at $D>1$ and $D=1$,
respectively) becomes even much more unpleasant after one starts
working in the formalism of SUSY QM. We reviewed briefly a few
standard attempts at a resolution of this problem (they did not
prove too satisfactory, anyhow) and showed that the analytic
continuation techniques mediated by the above-mentioned PTSY
mathematical formalism offer one of the best regularization
recipes. In particular, it enabled us to complete our older
results and to characterize the PTSUSY structure of the spiked
harmonic oscillator models {\em at any complex value} of the
strength $\alpha^2-1/4$ of their centrifugal-like spike.

Of course, the study of this subject is far from being completed.
We have outlined one of the eligible directions of the future
research by extending the scope of the present paper to the
exactly solvable angular equations. As we mentioned, their
importance lies in their indispensable role in the exact
solvability of several Hermitian models (viz., in the three-body
Calogero model and in the two-dimentional Smorodinsky-Winternitz
non-central oscillator etc). In such a setting we might emphasize
that within the broader framework of PTSY QM these angular
equations need not even remain exactly solvable at all.

A few further open question may be also formulated in the area of
a less direct overlap between SUSY QM and PTSY QM. As we
emphasized, the relativistic Klein-Gordon equation does not
provide a fully consistent description of a spinless particle but
still, it represents one of the phenomenologically most valuable
benchmark models in PTSY QM. In the two forthcoming appended
remarks, we intend to underline the existence of the new
mathematical challenges concerning the viability of its possible
transfer into the (possibly, regularized) formalism of SUSY QM.

  \subsection{Feshbach-Villars Hamiltonians and SUSY}

As we already mentioned, there exist several close formal
parallels between the Schr\"{o}dinger and Klein-Gordon equations
in the present regularization and analytic-continuation context.
The extent of these parallels is less clear when we start speaking
an algebraic language. For definiteness of a brief exposition of
this problem, let us assume that in a way paralleling the
constructions of the non-relativistic SUSY QM, the left (Hermitian
or non-Hermitian and regular or singular) Klein-Gordon operator is
factorizable, $H^{(KG)}_{(L)}= b \cdot a$. This means that we may
also factorize the two by two Feshbach-Villars matrix operator
postulating, for example, that
 \be
 H^{(FV)}_{(L)} = \left (
 \begin{array}{cc}
 0&H^{(KG)}_{(L)}\\
 I&0
 \ea
 \right )
 = B \cdot A, \ \ \ \
  A =\left [
 \begin{array}{cc} d&0\\ 0&a
 \ea
 \right ],
 \ \ \ \
 B=\left [
 \begin{array}{cc}
0& b^{}
\\
c&0 \ea \right ],
 \ \ \ \
 c \cdot d = I\,.
 \label{adsuperch}
 \ee
In a subtle way, such a postulate proves mathematically
inconsistent. Indeed, in the usual SUSY framework, operators
(\ref{adsuperch}) have to be inserted in the above-mentioned
general definition (\ref{superch}) of the supercharges ${\cal Q}$
and $\tilde{\cal Q}$, respectively, giving
 \be
 S_{SUSY,A} =
 \left [ \begin{array}{cccc}
 0&0&0&0\\
 0&0&0&0\\
 d&0&0&0\\
 0&a&0&0
 \ea
 \right ],\ \ \ \ \ \
 S_{SUSY,B} =
 \left [ \begin{array}{cccc}
 0&0&0&b\\
 0&0&c&0\\
 0&0&0&0\\
 0&0&0&0
 \ea
 \right ]
 \,
 \label{sssuperch}
 \ee
and obeying the same graded algebra as above. Routinely, the
partner Feshbach-Villars Hamiltonian becomes defined by the
formula
 \be
 H^{(FV)}_{(R)} = \left (
 \begin{array}{cc}
 0&H^{(KG)}_{(R)}\\
 I&0
 \ea
 \right )
 = A \cdot B, \ \ \ \
 H^{(KG)}_{(R)}= d \cdot b, \ \ \ \
 a \cdot c = I\,.
 \label{adveperch}
 \ee
This means that we need to find the quadruplet of operators
$a,b,c$ and $d$ where, in a suitable basis, the initial or
``input" factors $a$ and $b$ may be visualized as having the
respective annihilation- and creation-operator-type one-diagonal
and infinite-dimensional matrix forms,
 \be
 a=
 \left (
 \begin{array}{ccccc}
 0&a_0&0&\ldots& \\
 &0&a_1&0&\ldots \\
 &&&\ddots&
 \ea
 \right ), \ \ \ \ \ \
 b=
 \left (
 \begin{array}{cccc}
 0&\ldots&& \\
 b_0&0&\ldots &\\
 0&b_1&0&\ldots \\
 &&\ddots&
 \ea
 \right )\,.
 \ee
In this language the constraint $a\cdot c =I$ implies that in the
matrix
 \be
 c=\left (
 \begin{array}{ccccc}
 x_0&x_1&x_2&x_3&\ldots \\
 1/a_0&0&\ldots &&\\
 0&1/a_1&0&\ldots &\\
 0&0&1/a_2&0&\ldots \\
 &&&\ddots&
 \ea
 \right )
 \,
 \ee
the first row remains arbitrary. Unfortunately, this freedom is
not enough for the existence of a solution of our original
requirement $c\cdot d=I$. Indeed, even if we omit the upper-left
element of the latter matrix requirement, we get the unique
solution for this weaker constraint,
 \be
 d=
 \left (
 \begin{array}{ccccc}
 0&a_0&0&\ldots& \\
 &0&a_1&0&\ldots \\
 &&&\ddots&
 \ea
 \right ) \equiv a\,.
 \ee
{\it Vice versa}, we may evaluate the product
 \be
 c \dot d =\left (
 \begin{array}{ccccc}
 0&0&\ldots &&\\
 0&1&0&\ldots &\\
 0&0&1&0&\ldots \\
 &&&\ddots&
 \ea
 \right ) \equiv \Pi =\Pi^2 \neq I
 \,.
 \ee
This completes our constructive proof that the requirement $c\cdot
d=I$ has no solution at all. In the other words, the only
available Klein-Gordon-inspired super-Hamiltonian
 \be
 H_{KGSUSY}=
  \left [ \begin{array}{cccc}
 0&H^{(KG)}_{(L)}&0&0\\
 \Pi&0&0&0\\
 0&0&0&H^{(KG)}_{(R)}\\
 0&0&I&0
 \ea
 \right ],\ \ \ \ \ \
 H^{(KG)}_{(L,R)} = -\frac{d^2}{dx^2} + V^{(KG)}_{(L,R)}(x)\,.
 \label{dvmodel}
 \ee
connects merely the right Klein-Gordon field with an artificial,
``constrained" left SUSY partner which might only be interpreted
as a quasi-Klein-Gordon system equipped with an additional
projection-operator constraint $n_1^{(\pm)}=\Pi\,n_1^{(\pm)}$ at
all $n$. Its physical interpretation remains unclear.

  \subsection{Generalized time-evolution of $k-$th order}

When we compare the above Schr\"{o}dinger and Klein-Gordon SUSY QM
constructions, we may notice that they may be interpreted, quite
formally, as the first two elements of an infinite hierarchy of
the $k-$th order time-evolution equations
 \be
 \
 \left( i\,\p_t\right )^k \Psi^{(k)}(x,t)
 = H^{(k)}\,\Psi^{(k)}(x,t), \ \ \ \ \ \ \ \
 k = 1, 2, 3, \ldots\,.
 \label{jkafar}
 \ee
In a way which extends and parallels our previous notation we may
put
 \be
 \varphi_j= i\,\p_t\,\varphi_{j+1}, \ \ \ \ j = 1, 2, \ldots, k-1,
 \ \ \ \
 \varphi_k= \Psi^{(k)}(x,t)
 \ee
and arrange eq. (\ref{jkafar}) in the $k-$dimensional matrix form
 \be
 i\,
 \p_t
 \left (
 \ba
 \varphi_1\\
 \varphi_2\\
 \varphi_3\\
 \vdots\\
 \varphi_k
 \ea
 \right )
=
 \left (
 \begin{array}{ccccc}
 0&0&\ldots&0&H^{(k)}\\
 1&0&0&\ldots&0\\
 0&1&0&\ldots&0\\
 \vdots&&\ddots&&\vdots\\
 0&\ldots&0&1&0
 \ea
 \right )
 \cdot \left (
 \ba
 \varphi_1\\ \varphi_2\\
 \varphi_3\\
 \vdots\\
 \varphi_k
 \ea
 \right )\,.
 \label{jednaruka}
 \ee
This may be read as the new matrix problem $ i\,\p_t
\vec{\varphi}^{(k)} = {\cal H}^{(k)}\,\vec{\varphi}^{(k)}$ where
the matrix of the Hamiltonian obeys the pseudo-Hermiticity
condition
 \be
 \left [{\cal H}^{(k)} \right ]^\dagger = {\cal P}\,{\cal H}^{(k)}\,{\cal
 P}^{-1}, \ \ \ \ \ \ \ \
 {\cal P}=\left (
 \begin{array}{ccccc}
 0&0&\ldots&0&\eta\\
 0&\ldots&0&\eta&0\\
 \vdots&&_.\cdot^{^.}&&\vdots\\
 0&\eta&0&\ldots&0\\
 \eta&0&\ldots&0&0
 \ea
 \right )\,.
 \ee
The metric operators $\eta$ must be Hermitian and invertible and
may become positive or equal to unit operators in the respective
quasi-Hermitian or Hermitian special cases. In this sense the
Schr\"{o}dinger ($k=1$) and Klein-Gordon ($k=2$) evolution
equations may be viewed as special cases of a more general scheme
without, of course, any immediate applications within quantum
physics at $k \geq 3$.

\section*{Acknowledgement}

Work partially supported by the grant Nr. A 1048302 of GA AS CR.





\begin{thebibliography}{00}


\bibitem{CKS}
Cooper F,  Khare A and Sukhatme U 1995 Phys. Rep. 251 267

\bibitem{BB}
Bender C M and Boettcher S 1998 Phys. Rev. Lett.  24  5243;

Bender C M, Boettcher S and Meisinger P N 1999 J. Math. Phys. 40
2201

\bibitem{ostatni}
Caliceti E, Graffi S and Maioli M 1980 Commun. Math. Phys. 75 51;

Robnik M and Berry M V 1986 J. Phys. A: Math. Gen. 19 669;

Delabaere E and Pham F 1998 Phys. Letters A 250 25;

Fern\'andez F M, Guardiola R, Ros J and Znojil M 1998 J. Phys. {
A}: Math. Gen. 31 10105;

Mostafazadeh A 2003 J. Phys. A: Math. Gen. 36 7081;

Bender C M, Brody D C and Jones H F 2003 Am. J. Phys. 71 1095

\bibitem{Witten}
Witten E 1981 Nucl. Phys. B 188 513

\bibitem{onde}
Junker G 1996 Supersymmetric Methods in Quantum and Statistical
Physics (Berlin:Springer);

Bagchi B K 2001 Supersymmetry in Quantum and Classical Mechanics
(New York:Chapman and Hall);

Cooper F,Khare A and Sukhatme 2001 Supersymmetry in Quantum
Mechanics (Singapore:World Scientific).

\bibitem{tady}
cf. virtually all contributions to this special issue of  J. Phys.
A: Math. Gen.

\bibitem{BBprd}
Bender C M and Milton K A 1997 Phys. Rev. D 55 R3255;

Mostafazadeh A 1998 J. Math. Phys. 39 4499;

Bender C M and Milton K A 1999 J. Phys. A: Math. Gen. 32 L87

\bibitem{Levai}
Levai G and Znojil M 2001 Mod. Phys. Letters A 16  1973;

Levai G, Cannata and Ventura A 2002  Phys. Lett. A 300 271

\bibitem{Geyer}
Scholtz F G, Geyer H B and Hahne F J W 1992 Ann. Phys. (NY) 213 74

\bibitem{BBJ}
Bender C M, Brody D C and Jones H F 2002 Phys. Rev. Lett. 89
270401

\bibitem{BBJb}
Streater R F and Wightman A S 1964 PCT, spin and
statistics and all that (New York: Benjamin);

Mostafazadeh A 2003 J. Math. Phys. 44  974

\bibitem{Dirac}
Dirac P A M 1942 Proc. Roy. Soc. London A 180 1;

Pauli W 1043 Rev. Mod. Phys. 15 175;

Ramirez A and Mielnik B 2003 Rev. Mex. Fis. 49S2 130


\bibitem{PLB}
Znojil M, Cannata F, Bagchi B and Roychoudhury R 2000
         Phys. Lett. B 483 284;

Levai G and Znojil M 2002 J. Phys. A: Math. Gen. 35 8793;

Znojil M 2003 Rendiconti del Circ. Mat. di Palermo, Serie II,
Suppl. 71  199

\bibitem{hoptsusy}
Znojil M 2002 J. Phys. A: Math. Gen. 35 2341

\bibitem{NPB}
Znojil M 2003  Nucl. Phys. B 662 554

\bibitem{BG}
Buslaev V and Grecchi V 1993 J. Phys. A: Math. Gen. 26 5541

\bibitem{JR}
Jevicki A and Rodrigues J 1984 Phys. Lett. B 146 55

\bibitem{DP}
Das A and Pernice S 1999 Nucl. Phys. B 561 357

\bibitem{Ahmed}
Ahmed Z 2001 Phys. Lett. A 290 19

\bibitem{ptho}
Znojil M 1999 Phys. Lett. A 259  220

\bibitem{Mostafazadeh}
Mostafazadeh A 2002  J. Math. Phys. 43 2814

\bibitem{Herbst}
Herbst I 1999 private communication

\bibitem{AM}
Gohberg I, Lancaster P and Rodman L 1983 Matrices and Indefinite
Scalar Products (Basel: Birkh\"{a}user);

Mostafazadeh A 2002  J. Math. Phys. 43  6343 and erratum
Mostafazadeh A 2003  J. Math. Phys. 44 943

\bibitem{Andrianov}
Andrianov A A, Ioffe M V and Spiridonov V P 1993  Phys. Lett. A
174 273;

Andrianov A A, Ioffe M V, Cannata F and Dedonder J P 1995 Int. J.
Mod. Phys. A 10 2683;

Cannata F,  Junker G and Trost J 1998 Phys. Lett. { A 246} 219;

Klishevich  S M and Plyushchay M S 2002 Nucl. Phys. B 628 217;

Mostafazadeh A 2002 Nucl. Phys. B640 419;

Gunther U and Stefani F 2003 J. Math. Phys. 44 3097

\bibitem{Quesne}
Bagchi B, Mallik S and Quesne C 2002 Int. J. Mod. Phys. A 17 51

\bibitem{BW}
Bender C M and Wu T T 1969 Phys. Rev. 184 1231

\bibitem{sqw}
Bagchi B, Mallik S and Quesne C 2002 Mod. Phys. Lett. A17 1651

\bibitem{angulSE}
Znojil M 2003  J. Phys. A: Math. Gen. 36  7825

\bibitem{SmoWin}
Fri\v{s} I, Mandrosov V, Smorodinsky J, Uhl\'{\i}\v{r} M and
Winternitz P 1965 Phys. Lett. 16 354;

Evans N W 1991 J. Math. Phys. 32 3369;

Grosche C, Pogosyan G S and Sissakian A N 1995 Fortsch. Phys. 43
45

\bibitem{Calogero}
Calogero F 1969 J. Math. Phys. 10  2191;

Wojciechowski S R 1983 Phys. Lett. A 95 279

\bibitem{Jakubsky}
Jakubsk\'{y} V 2004 Czechosl. J. Phys. 54 67

\bibitem{Tater}
Znojil M and Tater M 2001 Phys. Lett. A 284 225

\bibitem{FV}
Feshbach H and Villars F 1958 Rev. Mod. Phys. 30 24

\bibitem{ego}
Znojil M 2004 Rendiconti del Circ. Mat. di Palermo, Ser. II,
Suppl. 72 211 (math-ph/0104012)

\bibitem{pseudo}
Mostafazadeh A 2003 Class. Quant. Grav. 20  155;

Mostafazadeh A 2004 Annals Phys. 309  1

\bibitem{CzW}
Znojil M 2004 Czechosl. J. Phys. 54 151

\end{thebibliography}
\end{document}